\documentstyle[12pt,epsfig]{article} 
\input{epsf.tex}
\newcommand{\One}{1\kern-4.5pt1}
\newcommand{\lapprox}{\raisebox{-0.5ex}{$\ 
\stackrel{\textstyle<}{\textstyle\sim}\ $}}
\begin{document}

\addtolength{\baselineskip}{0.20\baselineskip}


\hfill SWAT/98/193

\hfill July 1998


\vspace{48pt}

\centerline{\bf LATTICE APPROACH TO DIQUARK CONDENSATION }
\centerline{\bf IN DENSE MATTER} 


\vspace{18pt}

\centerline{\bf Simon Hands and Susan E. Morrison}
\centerline{\sl (UKQCD collaboration)}

\vspace{15pt}

\centerline{\sl Department of Physics, University of Wales Swansea,}
\centerline{\sl Singleton Park, Swansea SA2 8PP, U.K.}

\vspace{48pt}


\centerline{{\bf Abstract}}

\noindent
{\narrower 
We present results of a Monte Carlo simulation of a three dimensional 
Gross-Neveu
model with $\mbox{SU(2)}\otimes\mbox{SU(2)}$ chiral symmetry at non-zero baryon
chemical potential $\mu$, corresponding to non-zero baryon density. 
For $\mu$ sufficiently large there is a sharp
transition between a phase where the chiral symmetry is broken by a 
condensate $\langle\bar qq\rangle$ and one where 
a scalar diquark condensate $\vert\langle qq\rangle\vert\not=0$.
Global U(1) baryon number
symmetry may remain unbroken, however, due to the absence of long range
order in the phase of $\langle qq\rangle$. 
There is also tentative evidence for the
formation of a weaker pseudoscalar diquark condensate in the high density phase,
which violates parity.}




\vfill

\newpage

\section{Introduction}

The properties and behaviour of 
baryonic matter at high density have recently enjoyed renewed interest,
with the observation that the ground state even at densities
sufficiently high for chiral symmetry to be restored may be far from trivial
\cite{BL}\cite{ARW}\cite{RSSV}.
In brief, it is speculated that there is a regime of temperature and density 
for which the quarks have a large Fermi surface; kinetic considerations then
suggest that if there is any attraction at all between quarks, then the 
free-particle vacuum is unstable with respect to condensation of diquark pairs
from antipodal points on the surface, creating a gap between the Fermi energy 
and that of the lowest
excited states. 
This is precisely a relativistic incarnation of the BCS
instability in superconductors, with the diquarks playing the role of Cooper
pairs. 

The condensation 
phenomenon can be modelled by assuming the interaction between 
quarks is due to one-gluon exchange \cite{BL}, or using effective
four-fermion vertices resulting from the presence of instantons in the 
QCD vacuum \cite{ARW}\cite{RSSV}. A recent calculation using a
phenomenologically inspired Nambu -- Jona-Lasinio (NJL) model is given in 
\cite{BR}. Estimates of the gap energy are of order
100MeV, comparable with the chiral condensate at zero density and temperature.
It is intrinsically interesting in a field theoretic sense because
the diquark condensate may not be 
invariant under global U(1) rotations associated
with baryon number, and hence its formation 
be an example of spontaneous breaking of a
vectorlike symmetry. The condensate is also a gauge non-singlet, hence
promoting a dynamical Higgs effect, and making some (or perhaps all
\cite{ARW2}) of the gluons massive, leading to ``color superconductivity''. 
The pattern of symmetry breaking is predicted to be critically sensitive to the
number of light quark flavors present \cite{ARW2}.
Finally there may be interesting dynamical effects associated with the
competition between chiral $\langle\bar qq\rangle$ and diquark $\langle
qq\rangle$ condensates as the current quark mass $m$
and baryon chemical potential $\mu$ are varied. Phenomenologically, 
diquark condensation and its generalisations could have implications for 
the distribution of strangeness observed in heavy ion collisions \cite{W}, and
perhaps in the suppression of neutrino emission from neutron star cores,
resulting in slower cooling rates \cite{ARW}\cite{W}.

So far calculations have relied on model approximations to the QCD interaction, with ad hoc form
factors introduced, for instance, to model the effect of asymptotic freedom, 
which should suppress the condensate at large Fermi momentum, and hence 
large $\mu$. Many assumptions about the nature and origin of the $qq$
interaction are necessary. This is a natural area, therefore, for numerical
simulations of lattice QCD to make an impact. Recent simulations
of QCD in a fixed gauge have found evidence for weak diquark binding 
at zero density \cite{Frithjof}. Unfortunately, with $\mu\not=0$,
the Euclidean QCD path integral measure $\mbox{det}M(\mu)$,
where $M$ is the fermion kinetic operator, is complex, making standard
Monte Carlo simulation impossible. The most promising approach is to
locate the zeros of the QCD grand canonical partition function in the 
complex fugacity plane avoiding the the problem of the complex action
by generating the statistical ensemble using
$\det M(\mu = 0)e^{-S_G}$ (where $S_G$ is the gauge action) and reweighting 
with the factor $\frac{\det M(\mu\neq 0)}
{\det M(\mu= 0)}$ to achieve an overlap with the correct $\mu\neq0$ ensemble.
This method, though exact in principle, 
appears extremely slow to converge in practice, and the results, though 
promising, are far from definitive \cite{moreGlasgow}. To date the lattice
approach has only successfully been applied at non-zero density 
to toy field theories 
\cite{KKW}\cite{HKK}\cite{HKK2}.
In this paper we present the first lattice study of diquark condensation
in one such model, the Gross-Neveu (GN) model with
$\mbox{SU(2)}\otimes\mbox{SU(2)}$ chiral symmetry formulated in $d=2+1$
spacetime dimensions.

Which features of the GN model make it worth studying? In brief
\cite{RWP}\cite{HKK3}:
\begin{enumerate}

\item[$\bullet$] For sufficiently strong coupling the model exhibits
dynamical chiral symmetry breaking at zero temperature and density.

\item[$\bullet$] The spectrum of excitations contains both baryons 
(the elementary fermions), and mesons 
(composite fermion -- anti-fermion states),
which include a Goldstone mode.

\item[$\bullet$] For $2<d<4$ the model has an interacting continuum limit.

\item[$\bullet$] When formulated on a lattice, the model has a real Euclidean
action even for chemical potential $\mu\not=0$, and hence can be simulated
by standard Monte Carlo techniques.
\end{enumerate}
Therefore the model displays 
much of the essential physics
except for color confinement. Even such a 
simple theory may be useful in addressing some of the issues raised above, 
including the behaviour of competing condensates, 
and the full phase diagram in the $(\mu,T)$ plane once
temperature $T>0$. Simulations at $T=0$ \cite{HKK}\cite{HKK2}
have revealed a first order
chiral symmetry 
restoring transition at $\mu_c\simeq m_f$, where $m_f$ is the physical
(ie. constituent) fermion mass at $\mu=0$; this is in marked contrast to the 
pathological behaviour observed in approximations to QCD when the 
measure is held real, such as the quenched
approximation, where $\mu_c\sim m_\pi/2$ \cite{qQCD}. 

In the next section we will describe the lattice model in detail, and
identify which diquark operators we have studied using both lattice and
continuum notation. Our simulations and results are described in section 3, 
and a short discussion given 
in section 4.

\section{The Lattice Model}
\subsection{Action and Symmetries}

The lattice model we have simulated has the following Euclidean path
integral:
\begin{equation}
Z=\int D\sigma D\vec\pi\; \mbox{det}(M^\dagger M[\sigma,\vec\pi])\exp\left(
-{2\over g^2}(\sigma^2+\vec\pi.\vec\pi)\right),
\label{eq:Z}
\end{equation}
where $\sigma$ and the triplet $\vec\pi$ are real auxiliary
fields defined on the dual sites $\tilde x$
of a three dimensional Euclidean lattice. The determinant factor can be viewed
as having been obtained by integrating over Grassmann variables
$\chi,\bar\chi,\zeta,\bar\zeta$ with the following action:
\begin{equation}
S_{fer}=\bar\chi M[\sigma,\vec\pi]\chi + 
\bar\zeta M^*[\sigma,\vec\pi]\zeta
\end{equation}
with 
\begin{eqnarray}
M_{xy}^{pq}[\sigma,\vec\pi] & = & {1\over2}\delta^{pq}
\left[(\mbox{e}^\mu\delta_{yx+\hat0}-\mbox{e}^{-\mu}\delta_{yx-\hat0})
+\sum_{\nu=1,2}\eta_\nu(x)(\delta_{yx+\hat\nu}-\delta_{yx-\hat\nu})
+2m\delta_{xy}\right]\nonumber\\
& + & 
{1\over8}\delta_{xy}
\sum_{\langle\tilde x,x\rangle}\left(\sigma(\tilde x)\delta^{pq}
+i\varepsilon(x)\vec\pi(\tilde x).\vec\tau^{\,pq}\right).
\label{eq:M}
\end{eqnarray}
The parameters are the bare fermion mass $m$, the chemical potential
$\mu$ and the coupling $g^2$. The $\vec\tau$ are Pauli matrices
with indices $p,q=1,2$.
The symbols $\eta_\nu(x)$ denote the phases $(-1)^{x_0+\cdots+x_{\nu-1}}$, 
$\varepsilon(x)$ the phase $(-1)^{x_0+x_1+x_2}$,
and $\langle\tilde x,x\rangle$ the set of 8 
dual lattice sites neighbouring $x$. In this form we see that $\chi$
and $\zeta$ can be identified with isodoublet staggered lattice fermion fields
defined on the sites $x$. The integration measure in (\ref{eq:Z}) is manifestly
real\footnote{Note also that
$\mbox{det}M=\mbox{det}\tau_2M^*\tau_2$ is real \cite{MishaPC}}; 
note that $\chi$ and 
$\zeta$ have opposite-signed couplings to the fields $\pi_1$ and $\pi_3$.
It is straightforward to integrate over the 
auxiliary fields $\sigma$ and $\vec\pi$ to recover an action which contains 
fermions which self-interact via a four point interaction
$\propto g^2$; more details are
given for the corresponding four-dimensional model in \cite{HK}. 

In the limit $m\to0$, the model is invariant under a global symmetry
akin to the $\mbox{SU(2)}_L\otimes\mbox{SU(2)}_R$ of the continuum 
NJL model. Defining projection operators onto even and odd sublattices
${\cal P}_{e/o}$ by
\begin{equation}
{\cal P}_{e/o}(x)={1\over2}(1\pm\varepsilon(x)),
\end{equation}
we have
\begin{eqnarray}
\chi\mapsto({\cal P}_eU+{\cal P}_oV)\chi & ; & \bar\chi\mapsto\bar\chi
({\cal P}_eV^\dagger+{\cal P}_oU^\dagger)\nonumber\\
\zeta\mapsto({\cal P}_eV+{\cal P}_oU)\zeta & ; & \bar\zeta\mapsto\bar\zeta
({\cal P}_eU^\dagger+{\cal P}_oV^\dagger)\label{eq:SU2xSU2}\\
\Phi\equiv(\sigma\One+i\vec\pi.\vec\tau) & \mapsto & V\Phi U^\dagger\nonumber
\end{eqnarray}
with $U,V\in\mbox{SU(2)}$. Now for $T=\mu=0$, a mean field treatment
(which for a model with $N$ flavors of $\chi$ and $\zeta$
is equivalent to the leading order of an expansion in $1/N$)
shows that this $\mbox{SU(2)}\otimes\mbox{SU(2)}$ symmetry is
spontaneously broken to $\mbox{SU(2)}_{isospin}$ by the generation of a
condensate $\Sigma_0
=\langle\sigma\rangle={2\over g^2}\langle\bar\chi\chi\rangle$
and physical fermion mass $m_f=\Sigma_0$, the symmetry being broken for
coupling $g^2>g_c^2\simeq1.0$. Hence $\Sigma_0$ 
defines a physical scale in cutoff
units: a continuum limit exists as $g^2\to g_c^2$ from either phase. Quantum
corrections to this picture can be calculated as higher order terms in the $1/N$
expansion, which is renormalisable about $g^2=g_c^2$ \cite{RWP}\cite{HKK3}.
The numerical results in this
paper have been obtained for $N=1$ and a value 
$g^2=2.0$, corresponding to a zero density theory deep in the broken phase,
with $\Sigma_0=0.706(1)$ in units of inverse lattice spacing, ie.
rather far from the continuum limit.

We can also identify two other symmetries of (\ref{eq:Z},\ref{eq:M}), both of
which hold even for $m\not=0$. First, there is the global U(1) of baryon number:
\begin{equation}
\chi,\zeta\mapsto\mbox{e}^{i\alpha}\chi,\zeta\;\;\;;\;\;\;
\bar\chi,\bar\zeta\mapsto\mbox{e}^{-i\alpha}\bar\chi,\bar\zeta.
\label{eq:baryon}
\end{equation}
The chemical potential $\mu$ couples to the conserved charge associated with
this symmetry. Next, there is a discrete parity symmetry appropriate
to 2+1 dimensions:
\begin{eqnarray}
x=(x_0,x_1,x_2) &\mapsto& x^\prime=(x_0,1-x_1,x_2) \nonumber \\
\chi,\zeta(x)\mapsto(-1)^{x_1^\prime+x_2^\prime}\chi,\zeta(x^\prime)
&;&
\bar\chi,\bar\zeta(x)\mapsto(-1)^{x_1^\prime+x_2^\prime}\bar\chi,
\bar\zeta(x^\prime)
\label{eq:parity}\\
\sigma(\tilde x)\mapsto\sigma(\tilde x^\prime) &;&
\vec\pi(\tilde x)\mapsto-\vec\pi(\tilde x^\prime),\nonumber
\end{eqnarray}
whence the identification of $\sigma$ as scalar and $\pi$ as pseudoscalar.

For $T,\mu\not=0$ a mean field solution is also known \cite{RWP2}\cite{HKK},
in which $\Sigma(\mu,T)$ is expressed in terms of $\Sigma_0$. At zero
temperature the basic feature is that $\Sigma$ remains constant as $\mu$ is 
increased up to a critical value $\mu_c=\Sigma_0$, whereupon $\Sigma$ falls
sharply to zero (in the chiral limit), signalling a first order chiral symmetry
restoring transition. Monte Carlo simulations \cite{HKK} support
this picture, even when massless Goldstone excitations
(with the quantum numbers of the $\pi$ field) are present, and $N$ takes the 
minimal value $N=1$ \cite{HKK2}.

The reality of the measure in (\ref{eq:Z}) is an artifact of
our introducing conjugate flavors $\chi$ and $\zeta$. It is worth 
discussing how the Monte Carlo simulation manages to reproduce the 
main features of the expected behaviour of the model once $\mu\not=0$,
while similar simulations of QCD with quarks and conjugate quarks fail
dramatically \cite{qQCD}\cite{Misha}. 
In brief, it is because interactions between $\chi$ and $\zeta$
are negligible so that a light bound $\chi\zeta$
state, which would carry baryon number, does not form. The strongest
interactions are in the singlet channels $\chi\bar\chi$ and $\zeta\bar\zeta$,
which are dominated by disconnected ``bubble'' diagrams in the $1/N$ expansion,
and it is in these channels that Goldstone poles form \cite{LMHBK}.
Similarly, one might wonder how a diquark condensate, which is not invariant
under the vectorlike U(1) global symmetry 
of baryon number, could ever form in a model 
with a real measure, since such a breaking is usually forbidden by the
Vafa-Witten theorem \cite{VW}. The resolution is that the theorem does not hold
for models with Yukawa couplings to scalar degrees of freedom, such as the
coupling to $\sigma$ in (\ref{eq:M}). 

\subsection{Constructing Diquarks}

Next we discuss the operators used to form diquark pairs. We begin by expressing
the operators in terms of the lattice fields $\chi$ (we did not perform
measurements in the $\zeta$ sector, but the results here are identical).
In this study we have used four different local diquark operators: 
\begin{eqnarray}
NSS:\;\chi^p(x)\chi^p(x) & & 
NSP:\;\varepsilon(x)\chi^p(x)
\chi^p(x)\nonumber\\
SS:\;\chi^p(x)\tau_2^{pq}\chi^q(x)
&&SP:\;\varepsilon(x)\chi^p(x)
\tau_2^{pq}\chi^q(x).
\label{eq:latdiq}
\end{eqnarray}
Operators $NSS$ and $NSP$ 
are symmetric in both spatial and isospin indices, and so
would only form a non-zero condensate if extra flavors are introduced, ie.
$N>1$. In this case the operator could be written, with the indices $i,j$
running from 1 to $N$, ${1\over2}\chi_i\varepsilon_{ij}\chi_j$. For $N=1$ this
vanishes identically, but as described below, the diagram which would contribute
for $N\ge2$ can also be measured on an ensemble generated with $N=1$ in a
generalisation of the quenched approximation; we will refer to such operators as
{\sl non-spectral diquarks\/}. Operators $SS$ and $SP$, on the other hand, are
symmetric in space but antisymmetric in isospin, and hence can form even for
$N=1$; we will refer to these as {\sl spectral diquarks\/}. States
$NSS$ and $SS$ are even under parity (\ref{eq:parity}), and so will
be called {\sl scalar\/}, while $NSP$ and $SP$ are odd, and
hence {\sl pseudoscalar\/}. Note that 
$NSS$ and $NSP$ are not invariant under the chiral rotations (\ref{eq:SU2xSU2}),
or indeed under the remnant isospin symmetry in the chirally broken phase,
and all four
diquark operators not invariant under the U(1) of baryon number
(\ref{eq:baryon}). 

To gain further insight it is useful to transform to a basis in which
continuum-like spinor indices are shown. The recipe for staggered fermions in
three dimensions is well-known \cite{BB}. First we make a unitary transformation
to fields $u$ and $d$ defined on a lattice of twice the spacing of the
original, with site labels $y$:
\begin{equation}
u^{\alpha a}(y)={1\over{4\surd{2}}}\sum_A\Gamma^{\alpha a}_A\chi(A;y)\;\;\;;
\;\;\;d^{\alpha a}(y)={1\over{4\surd{2}}}\sum_A B^{\alpha a}_A\chi(A;y).
\end{equation}
The indices $\alpha,a$ each run from 1 to 2.
Here $A$ is a lattice vector with entries either 0 or 1, so that each site 
$x$ on the original lattice corresponds to a unique combination of $y$ and $A$.
The $2\times2$ matrices $\Gamma$ and $B$ are defined by
\begin{equation}
\Gamma_A=\tau_1^{A_0}\tau_2^{A_1}\tau_3^{A_2}\;\;\;;\;\;\;
B_A=(-\tau_1)^{A_0}(-\tau_2)^{A_1}(-\tau_3)^{A_2}.
\end{equation}
Now, with the definition 
\begin{equation}
q^{\alpha a}(y)=\left(\matrix{u^\alpha(y)\cr d^\alpha(y)\cr}\right)^a,
\end{equation}
we see that $q$ may be interpreted as a four-component spinor operated 
on by Dirac
matrices defined by:
\begin{equation}
\gamma_i=\left(\matrix{\tau_{i+1}& \cr&-\tau_{i+1}\cr}\right)\;\;
(i=0,1,2)\;;\;\;\;
\gamma_3=\left(\matrix{& -i\One\cr i\One&\cr}\right)\;\;\;;\;\;\;
\gamma_5=\left(\matrix{& \One\cr \One&\cr}\right),
\label{eq:gamma}
\end{equation}
with two 
flavors counted by the latin index $a$; this extra flavor degree of freedom is
due to the species doubling inherent in the lattice approach. It is now possible
to recast the fermion matrix (\ref{eq:M}) including the Yukawa interactions 
in continuum-like notation
\cite{BB}\cite{HKK3}. Here we will content ouselves with writing down
expressions for the diquark operators (\ref{eq:latdiq}) in the $q$-basis:
\begin{eqnarray}
NSS:\,-iq^T({\cal C}\gamma_5\otimes\tau_2\otimes\One)q && 
NSP:\,
-iq^T({\cal C}\otimes\tau_2\otimes\One)q\nonumber\\
SS:\,-iq^T({\cal C}\gamma_5\otimes\tau_2\otimes\tau_2)q
&&SP:\,-iq^T({\cal C}\otimes\tau_2\otimes\tau_2)q.
\label{eq:contdiq}
\end{eqnarray}
Here the first matrix in the direct product acts on the 4-spinor indices,
the second on the implicit flavor space indexed by $a$, and the third
on the explicit isospin degree of freedom indexed by $p,q$. The matrix 
${\cal C}$ is
the anti-unitary charge conjugation matrix defined by
${\cal C}\gamma_\mu {\cal C}^{-1}=-\gamma_\mu^*$; 
in our explicit basis (\ref{eq:gamma})
${\cal C}\equiv\gamma_0\gamma_2$. Denoting a symmetric state by $s$ and an
antisymmetric by $a$, we see operators $NSS$, $NSP$ are 
$a\otimes a\otimes s=s$, and $SS$ and $SP$ 
$a\otimes a\otimes a=a$, corroborating
our identification of spectral and non-spectral operators made earlier.

We can also rewrite the symmetries in the $q$-basis. The $\mbox{SU(2)}\otimes
\mbox{SU(2)}$ chiral symmetry (\ref{eq:SU2xSU2}) becomes
\begin{equation}
q_L\mapsto Uq_L\;\;\;;\;\;\;q_R\mapsto Vq_R\;\;\;\mbox{with}\;\;\;
q_{L/R}={1\over2}(1\pm\left(\gamma_5\otimes\One\otimes\One)\right)q,
\end{equation}
the U(1) of baryon number (\ref{eq:baryon}) 
\begin{equation}
q\mapsto\exp(i\alpha(\One\otimes\One\otimes\One))q\;\;\;;\;\;\;
\bar q\mapsto\bar q\exp(-i\alpha(\One\otimes\One\otimes\One)),
\end{equation}
while the parity transformation (\ref{eq:parity}) is now
\begin{equation}
q(x)\mapsto(\gamma_1\gamma_5\otimes\One\otimes\One)q(x^\prime)\;\;\;;\;\;\;
\bar q(x)\mapsto\bar q(x^\prime)(\gamma_5\gamma_1\otimes\One\otimes\One).
\end{equation}
The parity assignments of the diquark operators (\ref{eq:contdiq}) can be
checked by observing that $\gamma_1\gamma_5$ is a symmetric matrix.

\section{Simulations}

We have performed simulations of the model (\ref{eq:Z},\ref{eq:M})
on $L_s^2\times L_t$ lattices with bare mass $m=0.01$ at a coupling 
$1/g^2=0.5$, with $L_s$ varying between 8 and 24, and $L_t$ between 24 and 40.
The chemical
potential $\mu$ was varied between 0 and 1.2, 
with a critical value for chiral symmetry restoration $\mu_c\simeq0.65$.
The simulation method is a standard
hybrid Monte Carlo algorithm. We monitored the 
chiral condensate $\langle\bar\chi\chi\rangle$, and the baryon density 
$n$ defined by
\begin{equation}
n={1\over2}\left\langle\bar\chi^p(x)\mbox{e}^\mu\chi^p(x+\hat0)+
\bar\chi^p(x)\mbox{e}^{-\mu}\chi^p(x-\hat0)\right\rangle
\simeq\langle\bar q(\gamma_0\otimes\One\otimes\One)q\rangle
\end{equation}
using stochastic estimators.

To estimate diquark condensates we chose an indirect method: the two
point function $\langle qq(0)\bar q\bar q(x)\rangle$ 
was measured as the expectation
of the product of two fermion propagators as in QCD baryon spectroscopy, and
the condensate extracted by assuming clustering at large spacetime separation, 
ie.
\begin{equation}
\langle qq(0)\bar q\bar q(x)\rangle=
\langle qq(0)\bar q\bar q(x)\rangle_c+\langle qq\rangle\langle\bar q\bar
q\rangle,
\end{equation}
the latter term on the right being proportional to 
$\vert\langle qq\rangle\vert^2$. A non-zero condensate signal
therefore shows up as
a plateau in the timeslice correlator $G(t)$ at large time separation, and
may be extracted by a fit of the form
\begin{equation}
G(t)=\sum_{x_1,x_2}\langle qq(0)\bar q\bar q(t,x_1,x_2)\rangle
=A\exp(-M_+t)+B\exp(-M_-(L_t-t))+C^2(L_s)\vert\langle qq\rangle\vert^2,
\label{eq:timeslice}
\end{equation}
with $M_\pm$ the masses of forward and backward propagating diquark states 
respectively. The value of the constant $C(L_s)$ is not determined {\sl a
priori\/}. Naively one expects $C=L_s$, and hence the plateau height to 
be an extensive quantity.

We found 
by measuring the chiral condensate that chiral symmetry is restored 
for chemical potential greater than $\mu_c\simeq 0.65$.
The lattice 
virtually saturates with two $\chi$ quarks per site at $\mu=1.5$ corresponding
to a number density $n=2$. 

\begin{figure}
\vspace{-0.3cm}
\epsfig{file=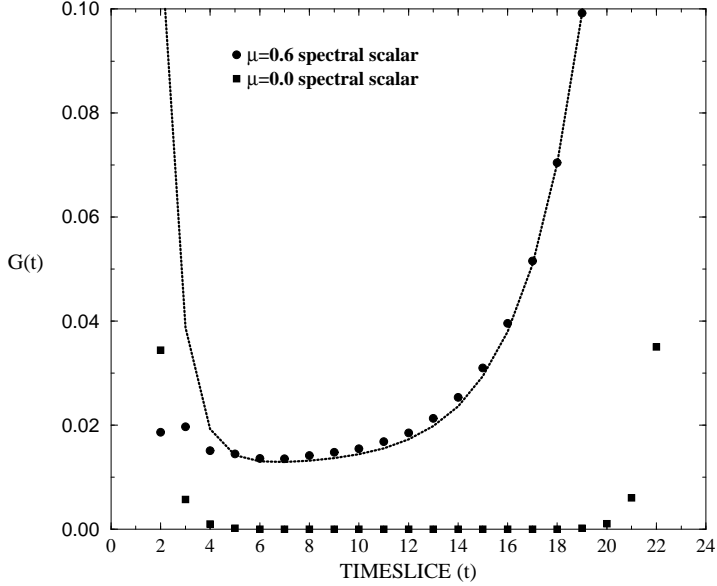,width=8cm,angle=-90,clip=}
\vspace{-0.2cm}
\caption{Spectral scalar correlators for two values of $\mu$ in the 
broken phase on $16^2\times 24$ lattices.}
\label{fig:G_broken}
\end{figure}
\begin{figure}
\vspace{-0.3cm}
\epsfig{file=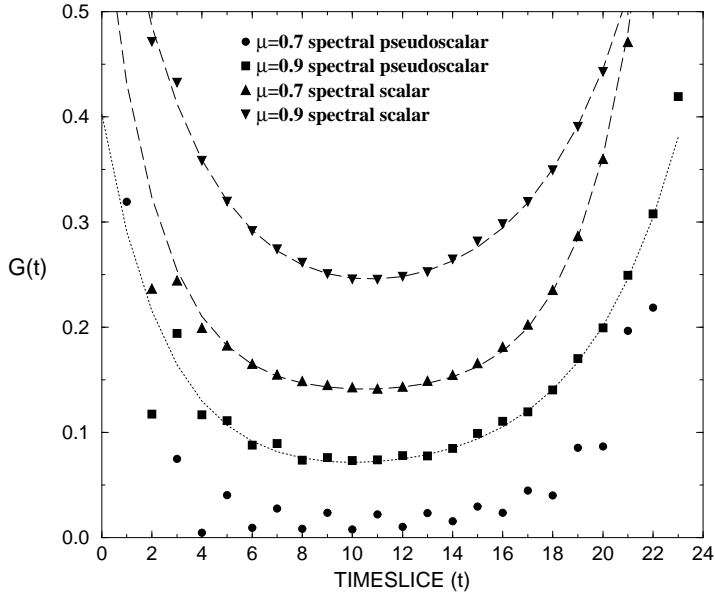,width=8cm,angle=-90,clip=}
\vspace{-0.2cm}
\caption{Spectral scalar and pseudoscalar correlators and their
fits for two $\mu$ values in the chirally symmetric phase
on $16^2\times 24$ lattices.} 
\label{fig:G_symmetric}
\end{figure}

The spectral scalar propagators in the broken phase $\mu<\mu_c$ are shown in 
Fig.~\ref{fig:G_broken}. For $\mu=0.0$ the spectral scalar condensate signal is
vanishing while for $\mu=0.6$ it is very small but non-vanishing. We found
no signal for the spectral pseudoscalar in the broken phase. The symmetric 
phase propagators and their fits are shown in Fig. \ref{fig:G_symmetric}. 
Noting the change
in scale on the $G(t)$ axis and comparing with Fig. \ref{fig:G_broken} we see
that in the chirally symmetric phase a strong signal develops for the spectral 
scalar which increases with chemical potential. The spectral pseudoscalar 
signal is weaker but distinctly non-zero and again increases with $\mu$. 
The non-spectral propagators are not shown here although
both the scalar and pseudoscalar closely follow the behaviour of the 
spectral pseudoscalar.

\begin{figure}
\vspace{-0.3cm}
\epsfig{file=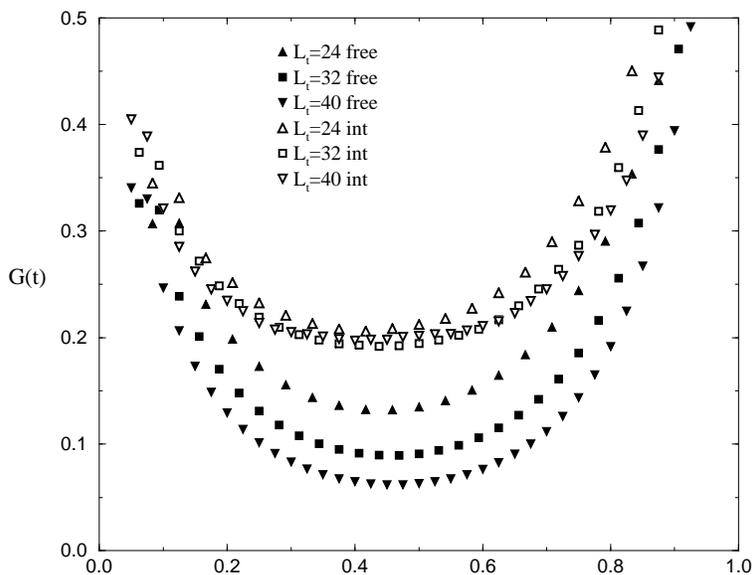,width=8cm,angle=-90,clip=}
\vspace{-0.2cm}
\caption{Spectral scalar correlators measured on $16^2\times L_t$
lattices at $\mu=0.8$, together with the corresponding results for 
free fermions. Note that the $t$ axis has been rescaled.} 
\label{fig:LTscal}
\end{figure}
The bulk of our results are from $16^{2}\times 24$
lattices. However, to assess possible systematic effects arising from
fitting the diquark propagators on lattices of finite temporal lattice 
extent $L_t$ we repeated simulations at $\mu=0.4$ 
on $16^2\times32$, and $\mu=0.8$ on $16^2\times32$ and
$16^{2}\times 40$ lattices. The results at $\mu=0.8$ are 
shown in Fig. \ref{fig:LTscal}, together with results obtained
for free fermions (ie. with $g^2=0$) on the same lattices with the same 
chemical potential. The time axis has been normalised
by $L_t^{-1}$ to aid comparison. Note that the plateau height 
in the middle of the lattice is almost constant as $L_t$ increases,
consistent with $\vert\langle qq\rangle\vert\not=0$, whereas in the free case 
the signal falls monotonically.
We found that the fits to the spectral scalar correlator 
in the symmetric phase ($\mu=0.8$) were very stable, yielding estimates for 
$C\vert\langle qq \vert\rangle$ of 0.425(1) for $L_t=24$, 0.424(1)
for $L_t=32$, and 0.436(1) for $L_t=40$. This is strong evidence
for a non-vanishing condensate.
The smaller parity violating signal, on the other hand, decreased by
about 25\% 
from 0.197(1) at $L_t=24$, via 0.167(1) at $L_t=32$, to 0.155(1) at $L_t=40$.
In the broken phase ($\mu=0.4$) the spectral scalar 
$C\vert\langle qq\rangle\vert$
was 0.065(1) for $L_t=24$ but consistent with zero on the $L_t=32$
lattice suggesting that the $\vert\langle qq\rangle\vert\neq 0$ 
result in the broken 
phase is at least in part due to finite lattice size. 

\begin{figure}
\vspace{-0.3cm}
\epsfig{file=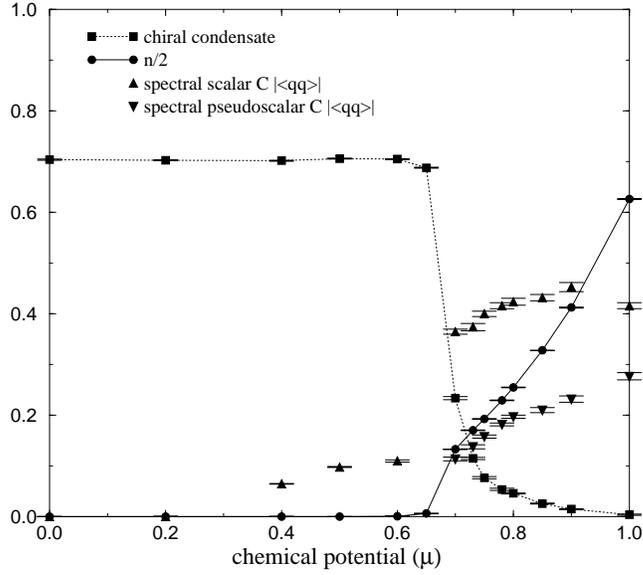,width=8cm,angle=-90,clip=}
\vspace{-0.2cm}
\caption{Overview of the observables $\langle\bar\chi\chi\rangle$, $n$ and
$SS$ and $SP$ diquark condensate signals
$C\vert\langle qq\rangle\vert$ as functions of $\mu$, obtained on a 
$16^2\times24$ lattice, showing the onset
of diquark condensation at $\mu_c\simeq0.65$.}
\label{fig:chir_transition}
\end{figure}

The trend in the diquark condensates as we pass from broken to symmetric phase
is clear from Fig. \ref{fig:chir_transition}. There is a critical value,
$\mu_c\simeq 0.65$, at which the chiral $\langle\bar{q}q\rangle$ condensate
falls sharply and the number density begins to rise from zero. 
The spectral scalar 
$C\vert\langle qq\rangle\vert$ 
condensate rises slowly from zero for $0.4 \le \mu<\mu_c$,
jumps upwards discontinuously close to $\mu_c$, and
continues to rise steadily. For
$\mu>1.0$ as the number density approached saturation the scalar 
$qq$ propagators became distorted and we were unable to 
achieve satisfactory fits.

The existence of 
a non-zero $\langle qq \rangle$ condensate in the broken phase 
could in principle convey physical information
about the onset 
of a nucleon liquid type phase. A non-zero 
condensate implies that 
there is a Fermi surface and therefore non-zero number density.
However, from the study with varying $L_t$ discussed above, it is 
clear that further study is required
to determine whether there is any non-zero signal 
associated with the onset of a nucleon 
liquid phase.

The spectral
pseudoscalar $C\vert\langle qq \rangle\vert$ signal was consistent with zero
for $\mu<\mu_c$ and no fits to the propagators were possible but for
$\mu>\mu_c$ this condensate is non-zero and increases with $\mu$. It is 
considerably smaller in magnitude than the spectral scalar. This pseudoscalar 
condensate is {\sl parity violating}, and
therefore it would be
remarkable if the 
signal in the chirally symmetric phase were to remain non-zero 
in the large volume limit.

\begin{figure}
\vspace{-0.3cm}
\epsfig{file=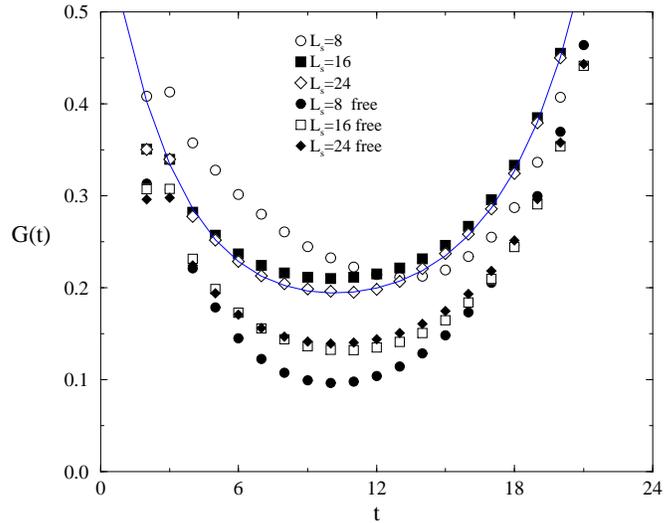,width=9cm,clip=}
\vspace{-0.2cm}
\caption{Spectral scalar diquark correlator 
at $\mu=0.8$ for both free and interacting quarks on $L_s^2\times24$ lattices.}
\label{fig:volume}
\end{figure}

To determine the behaviour with spatial volume we next performed a series of
runs on $L_s^2\times24$ lattices at $\mu=0.8$, with $L_s$ varying from 8 to
24. To our surprise we found very little change as $L_s$ increased, the
fitted value of $C\vert\langle qq\rangle\vert$ 
more or less saturating for $L_s\geq16$.
Results for $L_s=8,16,24$ are shown in Fig. \ref{fig:volume}, together with
the fit for $L_s=24$. For comparison we also plot
the equivalent correlators for free fermions at $\mu=0.8$; it is striking that
for this case the trend as $L_s$ increases is in the opposite 
direction. We conclude that the constant $C$ is roughly independent
of $L_s$; this puzzling behaviour will be further discussed in the final
section.

\begin{figure}
\vspace{-0.3cm}
\epsfig{file=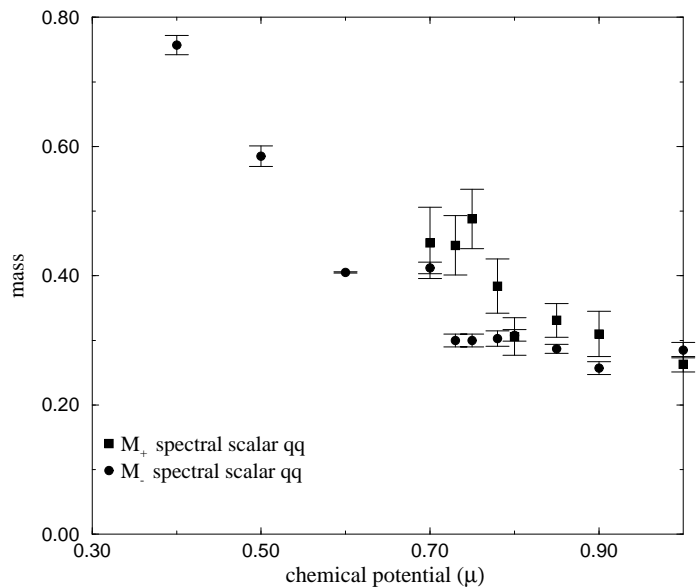,width=8cm,angle=-90,clip=}
\vspace{-0.2cm}
\caption{Masses of forward and backward moving states
$M_+$ and $M_-$ obtained from fits to the spectral scalar
propagators on a $16^2\times24$ lattice plotted as a function of $\mu$.}
\label{fig:masses}
\end{figure}
 
The diquark masses $M_+$ and $M_-$ obtained from the fits to the spectral scalar
$qq$ propagator are plotted as a function of $\mu$ in 
Fig. \ref{fig:masses}. We found that the forward
propagating $M_+$ states were more difficult to
extract from the fits than the backward propagating $M_-$ states, 
as reflected in the size of the error bars. The trend in the data is clear
with small masses in the symmetric phase and large masses in the broken phase.
The approximate linear decrease of $M_-$ with $\mu$ reflects a similar
trend observed in the physical fermion mass $m_f$ in previous
simulations \cite{HKK2}.
Notice that $M_-$ is approximately constant for $\mu>\mu_c$. The observed 
$qq$ states have masses around 0.3 which is light in
comparison to the zero density fermion mass $m_f=\Sigma_0\simeq0.7$,
which defines the ratio of physical to cutoff scales.
\begin{figure}
\vspace{-0.3cm}
\epsfig{file=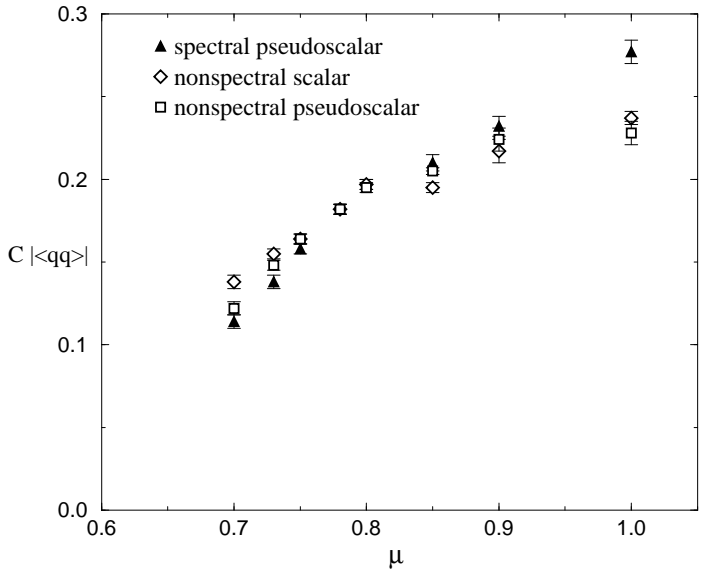,width=9cm}
\vspace{-0.2cm}
\caption{The three weaker diquark condensates $C\vert\langle qq\rangle\vert$
plotted in the high density phase
as a 
function of $\mu$.} 
\label{fig:qq_vs_mu}
\end{figure}

The spectral pseudoscalar and both nonspectral 
$C\vert\langle qq \rangle\vert$ condensates 
are plotted as a function of 
$\mu$ in Fig. \ref{fig:qq_vs_mu}. 
The signal for these three condensates
is much weaker than the signal for the spectral scalar condensate. 
The parity-violating spectral pseudoscalar and the two 
non-spectral condensates are almost identical in magnitude and have remarkably 
similar behaviour as a function of $\mu$. We also remark that we have observed
no evidence for spontaneous violation of isospin on inspection
of the fermion propagators.

To determine the effect of the bare quark mass on the $\langle qq \rangle$
condensates we performed simulations on $16^{2}\times 24$ lattices for
$m=0.02,0.05,0.08,0.1$ at $\mu=0.8$. The values of the condensates
were consistent within  errors with the $m=0.01$ result 
$C\vert\langle qq \rangle\vert$=0.424(7) for $m=0.02,0.05$ but the signal 
decreased by around 20\% to 0.342(10) for $m=0.08$ and 
by a further 20\% to 0.272(6) for $m=0.1$. In fact all of the diquark 
condensates decreased as we moved further from the chiral limit.  

\section{Summary and Outlook}

We have simulated the 3d GN model with $\mbox{SU(2)}\otimes\mbox{SU(2)}$ 
chiral symmetry at non-zero chemical potential
and found evidence suggestive of a diquark condensation, namely that
the two-point correlation function exhibits clustering at large temporal 
separation.
In order to quantify the measurement, however, it will be necessary to have a
numerical estimate for
the constant $C$ in (\ref{eq:timeslice}). 
We have up to now ignored the possible manifestations of fluctuations
in the phase of the condensate.
The spontaneous breaking of the U(1) symmetry (\ref{eq:baryon})
usually implies the existence of a Goldstone mode associated with long
wavelength fluctuations
in the phase of the $\langle qq\rangle$ condensate; this is not
accessible by the methods presented here, since $G(t)$ is real.
However, in the absence of an explicit diquark source (analogous
to a bare mass for the case of $\langle\bar qq\rangle$), one would expect
large finite volume effects, generically proportional to $L^{-(d^\prime-2)}$,
where $d^\prime$ is the dimension of the effective field theory
describing the fluctuations of the order parameter
\cite{Has}. We therefore speculate that the effective 
field theory describing the spatial correlations of the diquark
correlator has $d^\prime=d-1=2$, 
and that there is no long range order 
in the spatial directions \cite{Mermin}. A similar distinction between
temporal and spatial correlations has been observed in instanton liquid models
\cite{RSSV}.
This would account for the volume-independence of
$C$. Strictly speaking, therefore, we have not observed a true 
condensation. This interpretation of our results 
can be tested by simulations of the equivalent
$3+1$ dimensional model \cite{HK}, where we would expect long range order. 

In a BCS condensation, only quark pairs with momenta $\pm p$ close
to the Fermi surface contribute to the condensate. Therefore we expect
that $C^2$, which must depend on the number of participating 
$qq$ states, will be proportional to the area of the Fermi surface,
ie. $C^2\propto\mu^{d-2}$.
Note that the spectral scalar data in our work is well fitted by 
$C\vert\langle qq\rangle\vert\propto\surd\mu$ for $0.75\leq\mu\leq9$; fits
over a wider range of $\mu$, obtained from simulations closer to the 
continuum limit, would help to confirm this behaviour.
If we assume that $C\simeq O(\surd{\mu})$, 
we can claim to have detected
a strong signal for a spectral scalar $\vert\langle qq \rangle\vert$ condensate 
in the chirally
symmetric phase which is of the same order of magnitude as 
the chiral $\langle \bar{q}q \rangle$ condensate in the broken phase. 
This would imply that the
energy gap associated with diquark condensation is $O(\Sigma_0)$
and hence support the predictions of \cite{ARW}\cite{RSSV}\cite{BR}.

We have observed a signal for 
$\vert\langle qq \rangle\vert\neq 0$ in the broken phase possibly associated
with the onset of a nucleon liquid phase 
expected to occur for $\mu\lapprox\mu_c$, but further study of finite
size effects is needed for confirmation. 
We have also detected a weak signal for a parity violating 
$\vert\langle qq \rangle\vert$
condensate in the chirally symmetric phase. This may be a three
dimensional analogue of the axial diquark condensate reported in
\cite{ARW}. The remarks about
finite size systematics apply here too; however this fascinating phenomenon
could also be investigated by checking for the presence or absence of
parity doubling in the spectrum of spin-1 states. In a parity-symmetric 
vacuum we expect odd and even parity spin-1 states to be degenerate\footnote{
This was pointed out to us by M. Teper.}. This can be
observed in the spectroscopy of the 3d Thirring model \cite{thirr_long}.

 The two-point method used here to measure the $\langle qq \rangle$
 condensates requires the diquark
operators (\ref{eq:latdiq}) to be {\sl local\/} in the $\chi$ fields; 
this excludes
potentially interesting operators in this and related models.
The introduction of an explicit diquark source 
to study the one-point function directly is thus desirable,
both to calibrate our indirect two point function 
measurements, and 
to extend our study to other diquark operators and the $T\neq 0$
regime.

\section{Acknowledgements}

This work is supported
by the TMR-network ``Finite temperature phase transitions in particle physics''
EU-contract ERBFMRX-CT97-0122. Numerical work was done using an SGI Origin 2000
machine purchased by HEFCE for the UK Fundamental Physics Consortium.
We wish to thank the Center for Theoretical Physics, M.I.T., for 
hospitality while this work was being completed, and
have enjoyed stimulating discussions with Mark Alford,
Ian Barbour, Warren Perkins,
Krishna Rajagopal, 
Misha Stephanov and Frank Wilczek.


\begin{thebibliography}{xx}
%
\bibitem{BL} D. Bailin and A. Love, Phys. Rep. {\bf 107} (1984) 325.
\bibitem{ARW} M. Alford, K. Rajagopal and F. Wilczek, Phys. Lett. {\bf B422}
(1998) 247.
\bibitem{RSSV} R. Rapp, T. Sch\"afer, E.V. Shuryak and M. Velkovsky, 
hep-ph/9711396.
\bibitem{BR} J. Berges and K. Rajagopal, hep-ph/9804233.
\bibitem{ARW2} M. Alford, K. Rajagopal and F. Wilczek, hep-ph/9804403.
\bibitem{W} F. Wilczek, talk at the Bielefeld workshop {\sl QCD at Finite Baryon
Density\/}, hep-ph/9806395.
\bibitem{Frithjof} M. Hess, F. Karsch, E. Laermann and I. Wetzorke,
hep-lat/9804023.
\bibitem{moreGlasgow} I.M. Barbour, J.B. Kogut, S.E. Morrison, Nucl. Phys. 
(Proc. Suppl.) {\bf B53} (1997) 456; I.M. Barbour, S.E. Morrison, E.G. Klepfish,
J.B. Kogut and M.--P. Lombardo, Phys. Rev. {\bf D56} (1997) 7063;
Nucl.Phys. (Proc. Suppl.)
{\bf 60A} (1998) 220.
\bibitem{KKW} F. Karsch, J.B. Kogut and H.W. Wyld, Nucl. Phys. {\bf B280 [FS18]}
(1987) 289.
\bibitem{HKK} S.J. Hands, A. Koci\'c and J.B. Kogut, Nucl. Phys. {\bf B390}
(1993)
355.
\bibitem{HKK2} S.J. Hands, S. Kim and J.B. Kogut, Nucl. Phys. {\bf B442} 
(1995) 364.
\bibitem{RWP} B. Rosenstein, B.J. Warr and S.H. Park, Phys. Rep. {\bf 205}
(1991) 59.
\bibitem{HKK3}  S.J. Hands, A. Koci\'c and J.B. Kogut,
Ann. Phys. {\bf224} (1993) 29.
\bibitem{qQCD} I.M. Barbour, N.-E. Behilil, E. Dagotto, F. Karsch, A. Moreo,
M. Stone and H.W. Wyld, Nucl. Phys. {\bf B275 [FS17]} (1986) 296; 
C.T.H. Davies and
E.G. Klepfish, Phys. Lett. {\bf B256} (1991) 68;
J.B. Kogut, M.--P. Lombardo and D.K. Sinclair, Phys. Rev. {\bf D51} (1995) 1282;
Phys. Rev. {\bf D54} (1996) 2303; R. Aloisio, V. Azcoiti, G. DiCarlo, A.
Galante and A.F. Grillo, hep-lat/9804020.
\bibitem{MishaPC} M.A. Stephanov, private communication.
\bibitem{HK} S.J. Hands and J.B. Kogut, Nucl. Phys. {\bf B520} (1998) 382.
\bibitem{RWP2} B. Rosenstein, B.J. Warr and S.H. Park, Phys. Rev. {\bf D39} 
(1989) 3088.
\bibitem{Misha} A. Gocksch, Phys. Rev. {\bf D37} (1988) 1014, Phys. 
Rev. Lett. {\bf 61} (1988) 2054; M.A. Stephanov, Phys. Rev. Lett. {\bf76}
(1996) 4472.
\bibitem{LMHBK} M.-P. Lombardo, S.E. Morrison, S.J. Hands, I.M. Barbour and
J.B. Kogut, in preparation; S.E. Morrison, talk at the Bielefeld workshop
{\sl QCD at Finite Baryon Density\/}, hep-lat/9806033.
\bibitem{VW} C. Vafa and E. Witten, Nucl. Phys. {\bf B234} (1984) 173.
\bibitem{BB} C.J. Burden and A.N. Burkitt, Europhys. Lett. {\bf3} (1987) 545.
\bibitem{Has} 
P. Hasenfratz and H. Leutwyler, Nucl. Phys. {\bf B343} (1990) 241. 
\bibitem{Mermin} N.D. Mermin and H. Wagner, Phys. Rev. Lett. {\bf17} (1966)
1133.
\bibitem{thirr_long}L. Del Debbio, S.J. Hands and J.C. Mehegan, Nucl. 
Phys. {\bf B502} (1997) 269.
\end{thebibliography}
\end{document}